\documentclass[preprint,prc,aps]{revtex4}
\begin{document}

\title{Roper + Pion decays as a test of the radial resonances wavefunction}
                                                                                    
                                                                                    
\author{P.Stassart\footnote{Chercheur Qualifie FNRS ; Pierre.Stassart@ulg.ac.be}}
\affiliation{University of Liege, Fundamental Theoretical Physics, B5a, Sart Tilman, B-4000 Liege 1,Belgium}


\begin{abstract}
We calculate the decays of baryon resonances to the pion + Roper resonance
channel in the frame
of the flux-tube breaking model. The baryon and meson wavefunctions
we use are obtained from a linear confinement semi-relativistic
constituent quark model. The results, in good agreement with
the existing data, provide a spectrum independant test on the 
radial form of the main component of the Roper resonance. They shed light
on a likely orbital nature for the second delta resonance.
\end{abstract}

\maketitle


\section{Introduction}

The description of the baryon resonances spectrum with the use
of a one-gluon exchange potential as the source of splitting could not reproduce the ordering of the first two nucleon resonances, namely the N(1440) (Roper resonance) and the N(1520)~\cite{scal}.
While Goldstone Boson exchange models~\cite{glo} solve the ordering problem, they cannot account for the complete structure of baryon and meson spectra. It has been shown~\cite{rich} that the Hamiltonian used in nonrelativistic or semirelativistic quark models must yield a negative value for the Laplacian on at least part of the volume occupied by the resonance. Thus it is a likely possibility that models using both OGE and GBE potentials should be used in order to reproduce baryon spectra correctly.
Spectrum data, however, hardly allow to make a realistic tuning of the couplings of these interactions, as most splittings are small and of the same order of magnitude as the precision on them. Hence it may be interesting to look for other tests on the structure of the states. Decay amplitude calculations can be of most interest from this point of view. As the Roper Resonance is the most crucial resonance in studying these questions, it is most natural to think of calculating the Roper + pion decays of nonstrange baryon resonances. 
The direct data on the Roper + pion channel look scarse,~\cite{pdg}, as no lower bound to the decay widths to that channel is mentionned in the summary table besides the one of the radial partner of the roper, the $\Delta$(1600) resonance.
For this reason, that decay has not been calculated often. One example of this calculation is that of Capstick and Roberts~\cite{cap} who calculate the decay of the higher lying, N$>$2 band resonances. However, using the complete data contained in PDG, we can extract limits on the widths for all three and four star baryon resonances, including the N=1,2 bands. It is then worth calculating the Roper + pion decays of these states. We do this in the frame of the flux-tube model, using wavefunctions for the Roper resonance introduced in~\cite{imp}. In the next sections we describe the flux-tube model briefly, including the decay mechanism, and the way we treat the broad width of the Roper itself. Our results and discussion will be found in the last section. 

\section{The Flux Tube Model}

The long range attraction between quarks is modelized through the flux tube,
which corresponds to a linear three body confining potential. The spin-independant part of the Hamiltonian then reads
\begin{eqnarray}\label{eq:H0}
H_0=KE+{V(r_1,r_2,r_3)}+{E_0}.
\end{eqnarray}
 The first term stands for the (semirelativistic) kinetic energy, the last one stands for a constant, which can be set from the mass data. The potential consists of a short range part, which can be taken as the One Gluon Exchange potential, or as the Goldstone Boson Exchange potential, or any combination of them, and of a long range part that reads as the sum over all three quarks of the corresponding contributions
\begin{eqnarray}\label{eq:VLR}
V_i^{LR}=\kappa r_{i4}.
\end{eqnarray}
where $r_{i4}$ is the distance between the quark {i} and the point where the flux tubes meet at a $2\pi/3$ angle. For angles larger than this value that point is identical to the position of the corresponding quark. \\
In the case of the one gluon exchange potential, the short range part contains a scalar term, and spin dependant contributions, namely spin spin, tensor and spin orbit. The short range part reads as a sum over the quarks i,j = 1,2,3 with
\begin{eqnarray}\label{eq:VSR}
V_{ij}^{SR}=V_{scal}+V_{SS}+V_{T}+V_{SO}.
\end{eqnarray}
where $ V_{scal}=\frac{-4\alpha_S}{ 3 r_{ij}}$ and the spin dependent terms read as in ref.~\cite{imp}.
For each symmetry state, we use the corresponding SU(6) spin flavor wavefunction,
while the space part reads 
\begin{eqnarray}\label{eq:psin}
\psi_n(\rho,\lambda)=\phi_n(\rho,\lambda)\psi_0
\end{eqnarray}   
where $\rho$ and $\lambda$ are the Jacobi coordinates and $\phi_n(\rho,\lambda)$ ensure the orthogonality of the various wavefunctions.
For the radial excitation, which is essential in this process, we'll use~\cite{imp}
\begin{eqnarray}\label{eq:psirad}
\phi^*=N(1-\alpha(\rho^2+\lambda^2)^{\frac{k}{4}})
\end{eqnarray}
where N is a normalization factor, and $\alpha$ is set so as to insure the orthogonality with the ground state. The value k=4 corresponds to the harmonic oscillator case.
The hyperfine interaction is diagonalized in a truncated space corresponding to the first three levels of excitation. The resonances are then to be described by the sum of the products of the resulting mixing angles by the corresponding symmetry wavefunctions~\cite{imp}.

\section{decay widths calculation}

The decay mechanism we use is the breaking of an infinite flux tube, also called the $^3P_0$ quark pair creation (QPC) model~\cite{lopr}, which has been shown to be a good approximation of the finite extension QCD-inspired flux-tube breaking~\cite{ki,ss}.
The only parameter introduced by the model, the breaking amplitude $\gamma_0$,
is kept as the value yielding the $\Delta(1232)$ $->$ $N+\pi$ width~\cite{ss}.
To ensure we use the correct phase space volume, we use the experimental value for the mass of each resonance. This allows us to test wave function and mixing angles effects without interference due to specific spectral parameters. Moreover, this is essential in obtaining the correct phase space contribution for resonances below the $ N(1440) \pi$ threshold, as N(1520). 
Once the partial wave amplitudes $M^{J_R}_{ls}$are obtained ~\cite{ss,ss0}, they yield the partial widths $\Gamma_{ls}$ using
\begin{eqnarray}\label{gls}
\Gamma_{ls}= \frac{M^{J_B}_{ls}M^{*J_B}_{ls}k_\pi E_\pi E_R}{\pi(2J_B+1)M_B}
<I_RI_{\pi}I_{3R}I_{3\pi}/I_BI_{3B}>
\end{eqnarray}
where R,$\pi$,B stand for the Roper resonance, the emitted pion and the decaying baryon, $k$ for the momentum, $E$ for the recoil energy, $M_B$ for the baryon mass and the last factor is the Clebsch-Gordan coefficient in isospin space.
As the Roper resonance itself has a large width (350MeV), it is necessary to integrate the weighted value of the partial width over the Roper mass interval allowed for the decay process, namely from the Roper production threshold $M_N+M_\pi$ up to $M_B-M_\pi$. The weight $\sigma(M_R)$corresponds to the Relativistic Breit-Wigner distribution 
\begin{eqnarray}\label{weight}
\sigma(M_R)=\frac{2\Gamma}{\pi(\Gamma^2\frac{{(M_R^2-{M_{R0}}^2)}^2}{M_R^2})}
\end{eqnarray}
where $M_{R0}$ is the nominal value of the Roper resonance mass and $\Gamma$ the Roper width.     

\begin{table}{htb}
\caption{Square root of widths $(MeV)^{1/2}$}
\begin {tabular}{lccc}
\hline
  & {  Nominal mass  } 
  & {  Breit-Wigner  } 
  & {  PDG upper bound } \\
\hline
N(1520)D13 & 0 & 1.7 & 5.2\\
N(1675)D15 & 0.4 & 2.1 & 4.2\\
N(1680)F15 & 0.2 & 1.3 & 5.2\\
N(1700)D13 & 0.4 & 1.5 & 11.9\\
N(1710)P11 & 9.5 & 12.0 & 12.2\\
N(1720)P13 & 1.5 & 1.8 & 5.7 \\
$\Delta$ (1620)S31 & 8.6 & 8.0 & 8.8\\
$\Delta$ (1700)D33 & 1.8 & 2.3 & 8.9\\
$\Delta$ (1750)P31 & 2.1 & 2.4 & 16.5\\
$\Delta$ (1905)F35 & 0.7 & 1.1 & 12.4\\
$\Delta$ (1910)P31 & 17.6 & 14.9 & 15.1\\
$\Delta$ (1920)P33 & 6.9 & 6.0 & 10.0\\
$\Delta$ (1950)F37 & 2.4 & 3.5 & 12.6\\
\hline
\end{tabular}
\caption{Square root of widths $(MeV)^{1/2}$}
\label{tab:a}
\end{table}
\begin{table}
\begin{tabular}{lccc}
\hline
  & {  At nominal mass  }
  & {  BW weighted  }
  & {   PDG data} \\
\hline
k$=$1 & 2.8 & 11.5 & 8.5$^{+4.1}_{-3.5}$\\
k$=$2 & 5.9 & 27.2 \\
k$=$4 & 9.5 & 35.7 \\
\hline
\end{tabular}
\caption{Square root of widths$(MeV)^{1/2}$ for $\psi_R$ =$\psi_00$ (1-$\alpha (\rho^2 + \lambda^2))^{k/4}$}
\label{tab:b}
\end{table}

\section{Results and discussion}

In Table I, we display results for the square root of the width, that is the absolute value of the decays amplitudes (in MeV$^{1/2}$). These displayed here have been calculated using k=1 in the Roper wavefunction. No large variations are observed if one uses values of k up to 4, nor if one uses a pure radial symmetry for the Roper resonance. Thirteen resonances are listed, for which an upper bound can be extracted from PDG data. Resonances for which there exists a threshold effect due to the N$\eta$ channel, namely the S11 resonances, are not calculated here as the coupled channel calculation is beyond the scope of this article.
The comparison with data show that all model predictions fall below the experimantal upper bound. Although some of these bounds are rather large due to the lack of input data, the result is statistically significant. 
We notice that the effect of the integration with Breit-Wigner weight may be quite sizeable for resonance masses up to 1700 MeV, and is not negligible either for higher masses, as the width of the Roper is large enough for large variations in the amplitude to take place within the integration interval where the Breit-Wigner weight is not negligible.

Table II displays the results for the $\Delta$(1600) resonance, using k=1,2,4 in the radial content of both the Roper and $\Delta$(1600) wavefunctions, with the corresponding mixing angles.
One can notice that while the k=1 case yields a width value that lies within the experimental bounds, the k=2 and k=4 cases yield widths values far above the upper bound. It should be notice that while the content of the Roper resonance is rather insensitive to the wavefunction used (its radial content ranging from 91 to 96 percent of the total content),~\cite{imp,sas} this is not the case for its counterpart $\Delta$(1600). Indeed, while the harmonic limit (k=4) corresponds to an 81 percent radial content, the k=1 case corresponds to a largely orbital content.
We checked that a pure radial content yields a very large width, whatever the wavefunction used.
Hence, within the frame of the flux-tube model, we can affirm that the second $\Delta$ resonance should not be considered as a radial state, but rather as a mainly orbital excitation, with some coupling (mixing angle of order -0.175 ~\cite{imp}) to the radial symmetry state, while the true radial $\Delta$ partner of the Roper resonance is likely to be a higher mass, very large width resonance, with a very large coupling to the Roper $\pi$ channel.  
In conclusion, the model yields a good agreement with the experimental data for the N=1 and N=2 resonances decay to Roper $\pi$, for which, however, only large experimental intervals are available. The agreement is also good for the $\Delta$(1600) resonance when we use the form of the spatial radial wavefunction (k=1) corresponding to a higher short distance density~\cite{imp}. This sheds light on a likely orbital nature of the $\Delta$(1600) resonance. This indicates that any hyperfine splitting interaction should be compatible with an ordering placing the first radial nucleon resonance below the first negative parity orbital nucleon resonances, and placing the first radial delta resonance above the first positive parity orbital delta resonances altogether.




\end{document}